\documentclass[sigplan,10pt]{acmart}
\allowdisplaybreaks
\usepackage{microtype}
\AtBeginDocument{%
  }

\renewcommand\footnotetextcopyrightpermission[1]{}

\acmConference[EuroSys '27]{The 22nd European Conference on Computer Systems}{April 2027}{Rabat, Morocco}
\acmYear{2027}
\acmISBN{}
\acmDOI{}

\begin{document}

\title{MCF: A Memory- and Communication-Aware Co-Designed Framework for Long-Context LLM Inference}



\begin{abstract}
This paper presents Memory- and Communication-Aware Framework for Long-Context LLM Inference (MCF) that optimizes long-context inference through communication-aware and stage-adaptive KV-cache layout transitions. At the core of MCF is {Memory-Aware Replication and Scheduling} (MARS), a dynamic orchestration mechanism for chunked prefill. MARS exploits early-stage memory headroom by selectively replicating highly reused KV blocks, improving locality and hiding KV propagation latency behind attention computation. As prefill progresses and memory pressure increases, MARS progressively drains redundant replicas and converges the KV cache into a sharded decode-ready layout, avoiding expensive layout conversion at the prefill--decode boundary. For decoding, MCF adopts a query-centric execution model in which KV shards remain stationary while lightweight query tensors are routed across workers. This changes communication from KV-cache-dependent transfer to query-dependent transfer, reducing communication growth with context length. MCF further integrates a hardware-software co-designed execution layer based on an FPGA compute-NIC datapath, enabling metadata-driven KV routing and collective operations directly in the network path. Experiments show that MCF improves TTFT, throughput, and scalability over state-of-the-art distributed inference systems, with larger benefits as context length and cluster size increase. 
\end{abstract}

\keywords{LLM Serving, Distributed Inference, KV-Cache Management, In-Network Computing, Zero-copy Communication}

\author{Huaizhi Zhang}
\affiliation{%
  \institution{University of Essex}
  \country{United Kingdom}
}
\email{hz24245@essex.ac.uk}

\author{Jiacheng Zhu}
\affiliation{%
  \institution{University of Essex}
  \country{United Kingdom}
}
\email{jz23222@essex.ac.uk}

\author{Xuqi Zhu}
\affiliation{%
  \institution{University of Essex}
  \country{United Kingdom}
}
\email{xz18173@essex.ac.uk}

\author{Junkyu Lee}
\affiliation{%
  \institution{University of Essex}
  \country{United Kingdom}
}
\email{j.lee@essex.ac.uk}

\author{Klaus McDonald-Maier}
\affiliation{%
  \institution{University of Essex}
  \country{United Kingdom}
}
\email{kdm@essex.ac.uk}

\author{Xiaojun Zhai}
\affiliation{%
  \institution{University of Essex}
  \country{United Kingdom}
}
\email{xzhai@essex.ac.uk}

\renewcommand{\shortauthors}{Zhang et al.}

\maketitle

\section{Introduction}
Long-context inference is becoming a critical workload for Large Language Models (LLMs). Applications such as retrieval augmented generation~\cite{lewis2020retrieval}, code generation~\cite{liu2024repobench}, document analysis~\cite{ma2024mmlongbenchdoc}, and agentic reasoning often require prompts with tens of thousands of tokens. 
For such workloads, the prefill stage becomes a dominant contributor to Time-To-First-Token (TTFT), because the system must process the entire input context and materialize the corresponding Key-Value (KV) cache before decoding can begin.

Many existing distributed inference systems rely on tensor parallelism and pipeline parallelism. Tensor parallelism partitions model weights across devices, while pipeline parallelism partitions model layers. These techniques are effective for scaling model parameters and computation, but they do not directly address the sequence-dependent KV-cache growth in long-context inference. During prefill, each new query chunk must attend to all historical KV blocks. During decode, every generated token must access the increased KV cache. As a result, long-context inference introduces different types of memory and communication bottleneck to conventional model-parallel weight partitioning.

Context parallelism addresses this issue by partitioning the sequence dimension and distributing KV cache across workers. However, existing context-parallel designs~\cite{li2023sequenceparallelism, liu2024ringattention} often rely on static KV sharding and fixed communication patterns, such as ring-style KV propagation. While such designs preserve a memory-efficient layout, they repeatedly move historical KV blocks during inference. As the number of chunks grows, this repeated communication becomes increasingly expensive under bandwidth-constrained distributed inference. It would also be inefficient in heterogeneous or dynamically changing communication environments, where link bandwidth, latency, or node availability can vary over time.

Moreover, we observe that prefill is not a uniform execution phase. Early in prefill, the accumulated KV cache is small and each worker has substantial memory headroom. At the same time, early KV blocks have the highest future reuse because they will be attended by many subsequent chunks. Near the end of prefill, memory pressure increases and the system must recover a compact KV layout suitable for decode. This temporal asymmetry suggests that a single static KV-cache layout is suboptimal.

We thus propose MCF, a \underline{M}emory-\underline{C}ommunication-Aware Distributed Inference \underline{F}ramework for Long-Context LLMs to response to this challenge. At the core of MCF is MARS, a stage-aware KV-cache orchestrator that dynamically controls KV replication and convergence during chunked prefill. MARS temporarily replicates high-reuse KV blocks when memory headroom is available, reduces the frequency of remote KV accesses and the critical-path latency of KV transfers in later chunks, and progressively retires redundant replicas as their reuse window expires. By the end of prefill, MARS guarantees that all KV blocks converge to an owner-only sharded layout, enabling efficient KV-stationary decode.

MCF further introduces a distributed chunked-prefill execution model that pipelines local computation with KV propagation and missing-block fetching. Rather than treating communication as a serialized phase after computation, MCF overlaps chunk execution with dynamic KV movement. This improves TTFT for long-context inference.

To realize this dynamic execution efficiently, we design a hardware-software co-designed unified execution framework. The software runtime emits MARS metadata, including routing identifiers, operation codes, KV block identifiers, and destination sets. The hardware maps this metadata to an FPGA compute-NIC datapath that performs KV propagation, collective communication, and decode-state reduction in the data plane. This design translates MARS's dynamic scheduling decisions into high-performance distributed inference.

This paper makes three key contributions:

\begin{itemize}
    \item We propose \textbf{MARS}, a stage-aware KV-cache orchestrator that dynamically replicates, drains, and converges KV blocks across prefill stages, exploiting early memory headroom while guaranteeing a sharded decode-ready layout.

    \item We design a \textbf{distributed chunked-prefill execution model} that pipelines local computation with KV propagation and remote KV-cache fetching, improving compute-communication overlap and reducing TTFT for long-context inference.

    \item We design a \textbf{hardware-software co-designed unified execution framework} that bridges MARS and hardware execution. By integrating a metadata-driven FPGA compute-NIC datapath, the framework realizes dynamic KV routing and collective operations in hardware, enabling high-performance distributed inference.
\end{itemize}

\section{Background and Motivation}
In this section, we first discuss distributed LLM inference from the perspective of KV-cache memory requirements. We then briefly review tensor parallelism, pipeline parallelism, sequence parallelism, and context parallelism. Finally, we identify the key bottlenecks in existing long-context LLM inference systems, particularly those arising from static KV-cache layouts, and use these limitations to motivate our work.

\subsection{Distributed LLM Inference}

Large language model inference is commonly divided into two stages: prefill and decode. Given an input prompt of length $S$, the prefill stage processes all prompt tokens and materializes KV cache for every transformer layer. The decode stage then generates output tokens autoregressively, where each new query attends to the accumulated KV cache.

As context length increases, the KV cache becomes a major memory and communication bottleneck. For a model with $L$ transformer layers, $H_{kv}$ KV heads, head dimension $d$, and element size $b$, the KV-cache memory per token is
\begin{equation}
M_{\mathrm{token}} = 2LH_{kv}db
\label{eq:m-token}
\end{equation}
where the factor of two corresponds to keys and values. Therefore, the KV-cache footprint grows linearly with sequence length. Long-context inference further amplifies this issue because each prefill chunk must attend to all historical KV blocks, while every decode step must access the full accumulated KV cache. This makes long-context inference fundamentally different from short-context inference: the system is no longer dominated only by model weights and matrix multiplication, but also by the placement, movement, and reuse of KV-cache states.

\subsection{Tensor Parallelism}

Tensor parallelism (TP) is a widely used method for scaling large transformer models across multiple accelerators. TP partitions large matrix operations within each transformer layer across devices. For example, linear projections in attention and feed-forward networks can be split along column or row dimensions, and devices exchange partial results through collective communication such as all-reduce or all-gather. This approach is effective for distributing model parameters and compute within a layer, and has been adopted in systems such as Megatron-LM~\cite{shoeybi2019megatron}.


\subsection{Pipeline Parallelism}

Pipeline parallelism (PP) partitions a transformer model along the layer dimension. Each device stores and executes a subset of layers, and microbatches are pipelined across stages. PP is effective for distributing model weights across devices and increasing throughput for batched workloads. It has been combined with tensor and data parallelism in large-scale systems such as Megatron-LM~\cite{narayanan2021efficient}.


\subsection{Sequence Parallelism}

Sequence parallelism (SP) partitions tensors along the sequence dimension. Instead of replicating the entire sequence on every device, SP distributes sequence tokens or activations across workers. This can reduce activation memory and expose additional parallelism for long-sequence transformer computation. Megatron-style sequence parallelism has been used to reduce activation memory overhead in large transformer training~\cite{korthikanti2023reducing}. More recent long-context systems further explore SP as a way to scale sequence length beyond the memory capacity of a single accelerator.


\subsection{Context Parallelism}

Context parallelism (CP) can be viewed as a sequence-parallel strategy specialized for long-context attention and KV-cache management. In CP, the input context and KV cache are partitioned across devices. Each worker owns a subset of the sequence and participates in distributed attention computation. This layout is attractive because it reduces per-device KV-cache memory from $O(S)$ to approximately $O(S/N)$ for $N$ workers. Existing context parallelism methods mainly differ in how they communicate attention states across devices. Two representative approaches are Ring Attention\cite{liu2024ringattention} and DeepSpeed-Ulysses\cite{jacobs2023deepspeed}. 

\paragraph{Ring Attention.} Ring Attention partitions the sequence into blocks and arranges workers in a logical ring. During attention computation, each worker computes partial attention using local KV blocks while KV blocks are circulated around the ring. This blockwise design enables overlap between KV communication and attention computation, and allows long sequences to be processed across multiple devices~\cite{liu2024ringattention}. Ring Attention is memory efficient because each KV block can remain owned by one worker. However, during chunked prefill, historical KV blocks may be repeatedly streamed across workers for later chunks. As the length of input sequence increases, the total KV movement can grow rapidly.

If the prompt is divided into $C$ chunks and each chunk attends to all previous chunks, a ring-style design may communicate prefix KV blocks repeatedly. The cumulative number of prefix-block accesses grows as
\begin{equation}
\sum_{c=1}^{C} c = \frac{C(C+1)}{2} = O(C^2)
\end{equation}

This repeated static communication can substantially increase TTFT during long-context prefill, particularly under bandwidth-constrained interconnects, when KV-transfer latency cannot be fully hidden behind block-level attention computation.


\paragraph{DeepSpeed-Ulysses.}
DeepSpeed-Ulysses is another long-sequence parallelism approach that partitions the sequence dimension and uses collective communication to redistribute tensors for attention computation~\cite{jacobs2023deepspeed}. Ulysses uses all-to-all communication to transform distributed tensor layouts so that attention heads or sequence partitions can be processed efficiently. Compared with naive sequence parallelism, Ulysses improves scalability for long-sequence training by carefully designing communication and layout transformations.

However, Ulysses-style all-to-all communication still relies on relatively structured collective operations. For long-context inference, especially chunked prefill followed by decode, the optimal communication pattern can change over time. Early in prefill, memory pressure is low and historical KV blocks have high future reuse. Later in prefill, memory pressure increases and the system must converge to a compact sharded KV layout for decode. A fixed communication/layout strategy cannot fully exploit this temporal structure.

\subsection{Limitations of Static KV Layouts}

The above approaches demonstrate that distributing the sequence dimension is necessary for long-context inference. However, most existing designs rely on relatively static execution layouts. Ring Attention keeps an owner-only KV layout and repeatedly propagates KV blocks. Ulysses-style systems rely on structured all-to-all layout transformations. These approaches are effective for regular distributed attention computation, but they do not explicitly optimize the temporal reuse pattern of chunked prefill.

The key observation behind MCF is that prefill is not a uniform phase. At the beginning of prefill, only a small number of KV blocks have been generated, so each worker has memory headroom. At the same time, early KV blocks have the highest future reuse because they will be attended by many subsequent chunks. Near the end of prefill, memory pressure becomes high, and the system must recover a balanced sharded KV layout for decode.

This temporal asymmetry suggests that neither pure sharding nor full replication is optimal. Pure sharding minimizes memory usage but can repeatedly fetch or stream historical KV blocks. Full replication reduces remote KV access but is not memory scalable and leaves an inefficient layout for decode. The desired design should temporarily use available memory to replicate high-reuse KV blocks during early prefill, then progressively retire redundant replicas and converge to a sharded layout before decode.
\subsection{Motivation for MCF}
MCF is motivated by the need for a dynamic, stage-aware execution layout for long-context inference. Instead of treating KV placement as a fixed policy, MCF treats it as a runtime scheduling problem. Its core mechanism, MARS, divides prefill into multiple stages and dynamically orchestrates KV-cache replication, draining, and convergence according to memory headroom and expected future reuse.

During early prefill, MARS(Section~\ref{sec:mars}) creates temporary replicas for high-reuse KV blocks to reduce repeated remote KV access. During middle prefill, MARS adjusts replica factors as memory pressure increases. By the end of prefill, MARS guarantees that each KV block converges to its owner-only shard, producing the same decode-ready layout as conventional context-parallel inference.

MCF further combines this dynamic software policy with a hardware-software co-designed execution substrate. Runtime scheduling decisions are encoded as compact metadata describing routing, operation, and tensor layout information. The FPGA compute-NIC datapath then uses this metadata to perform KV dissemination, collective communication and decode-state reduction.

\section{MCF Architecture Overview}

MCF is a distributed long-context LLM inference framework that coordinates KV-cache placement, communication, and hardware execution across prefill and decode. Its central design principle is that the preferred KV layout is stage-dependent. During chunked prefill, early KV blocks are reused by many later chunks, while late KV blocks have limited remaining reuse. MCF therefore allows temporary replication of high-reuse KV blocks when memory headroom is available. Before decode begins, these temporary replicas are progressively drained, leaving a balanced owner-only sharded KV layout.

\begin{figure}[htbp]
    \centering
    \includegraphics[width=\linewidth]{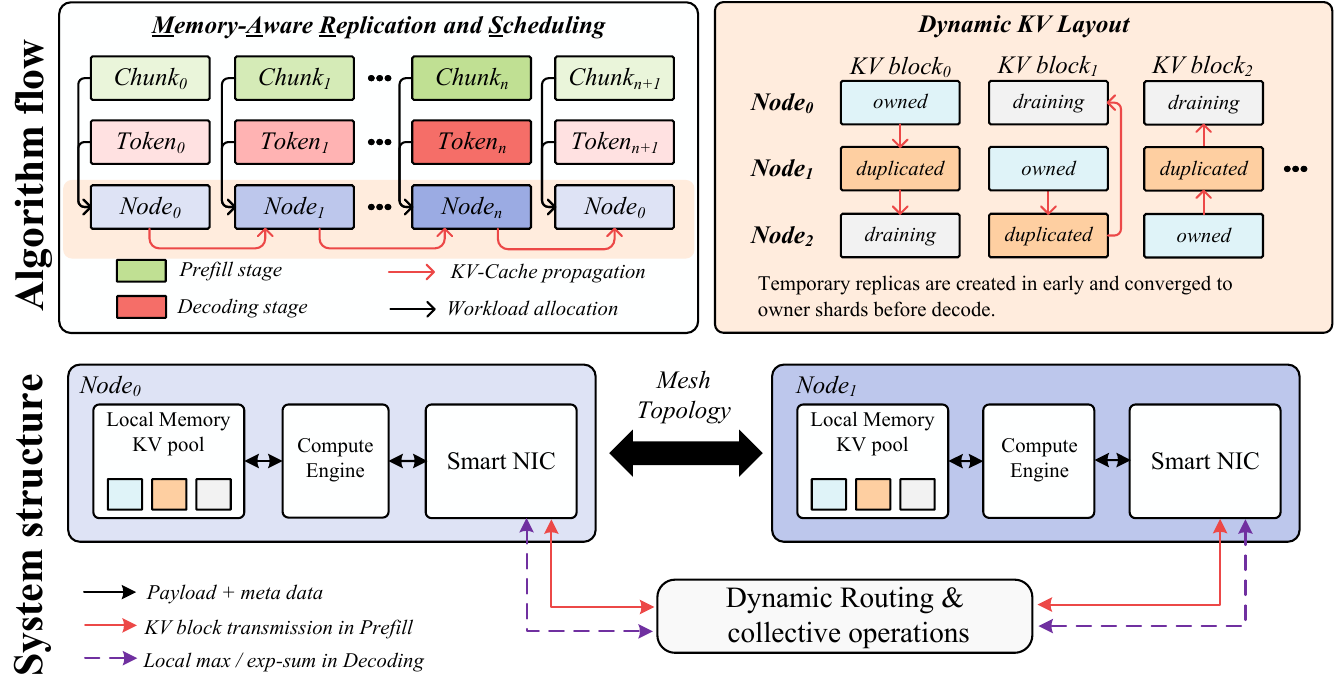}
    \caption{MCF Overview: MCF dynamically schedule the KV cache propagation and replication to balance the communication cost and memory overhead in context parallelism by using Memory-Aware Replication and Scheduling (MARS).}
    \label{fig:Overview}
\end{figure}

Figure~\ref{fig:Overview} shows the overall design of MCF. The framework consists of two components. The first is MARS, a runtime scheduler that determines the replica factor, owner node, and routing decision for each KV block according to memory capacity, expected reuse, and communication cost. The second is a metadata-driven FPGA compute-NIC datapath that executes these decisions by coupling local transformer computation with in-network KV routing and state reduction.

\subsection{Progressive KV Layout}
\label{subsec:progressive_kv_layout}

MCF processes the input prompt in chunks. For each generated KV block, MARS assigns a permanent owner node according to the final sharding policy. The owner copy is never removed. In addition, MARS may create temporary replicas during prefill to reduce later remote KV accesses. These replicas are most beneficial for early KV blocks because they have the largest future reuse window.

As prefill progresses, accumulated KV-cache memory increases and the benefit of replicating newly generated blocks decreases. MARS therefore gradually reduces replica factors and marks redundant replicas for draining. Draining replicas may still serve in-flight computation, but they are excluded from future scheduling decisions and are removed once no longer needed. This process avoids bulk layout conversion: because the owner copy always remains valid, convergence only requires retiring temporary replicas rather than migrating KV data.

At the end of prefill, all temporary replicas are removed and each KV block remains only on its owner node. This owner-only layout is decode-ready. During decode, MCF keeps KV shards stationary and routes only the current query or compact partial attention states across nodes. Each node computes attention over its local KV shard, and the partial results are reduced to form the global attention output.

\subsection{Compute-NIC Execution Substrate}
\label{subsec:compute_nic_execution_substrate}

To execute this progressive layout efficiently, each MCF node integrates a local compute engine, a KV memory pool, and a metadata-driven FPGA compute-NIC datapath. The compute engine performs local transformer computation and emits either KV blocks during prefill or compact partial attention states during decode. The KV memory pool stores both permanent owner blocks and temporary replicas.

The MARS runtime attaches compact metadata to each tensor payload, including the execution stage, operation type, KV-block identifier, route identifier, destination set, and reduction operator. The FPGA NIC datapath interprets this metadata to perform the corresponding communication action, such as KV propagation during prefill or partial-state reduction during decode. By moving routing and collective operations into the datapath, MCF avoids host-side communication control on the critical path and enables computation and communication to proceed in a pipelined manner.

Overall, MCF uses different communication policies for different inference stages: prefill uses memory-aware temporary KV replication to reduce exposed remote access latency, while decode uses a KV-stationary sharded layout to avoid moving accumulated KV cache. This progressive transition is the basis for the MARS scheduling algorithm described in the following section.

\section{MARS: Memory-Aware Replication and Scheduling}
\label{sec:mars}

MCF relies on dynamic KV-cache layouts whose effectiveness depends on choosing appropriate runtime parameters, including the prefill chunk size, KV block granularity, replica factor, and replica convergence schedule. A naive configuration can either under-utilize the interconnect, causing pipeline stalls, or over-replicate KV blocks, increasing memory pressure and unnecessary communication. To guide these decisions, we introduce MARS, a Memory-Aware Replication and Scheduling algorithm that jointly considers memory capacity, compute throughput, interconnect bandwidth, and future KV reuse. The goal of MARS is to minimize time-to-first-token (TTFT) under bounded KV-cache memory capacity while ensuring that all temporary replicas eventually converge to a decode-friendly sharded KV layout.

We consider a distributed inference system with $N$ worker nodes. Each node — whose detailed design we defer to Section \ref{sec:node} — composed of an accelerator for transformer computation and a communication engine for KV-cache routing. Given an input prompt of $S$ tokens, MCF performs chunked prefill with chunk size $B$, producing $C=S/B$ prefill chunks. For a transformer-based model with $L$ layers, $H_q$ query heads, $H_{kv}$ KV heads (e.g., in group query attention~\cite{ainslie2023gqa}, $H_{q} \neq H_{kv}$), head dimension $d$, and element size $b$ bytes, the KV-cache footprint per token is $M_{\mathrm{token}}$ (Equation~\ref{eq:m-token}).
Therefore, the bytes required for KV-cache generated by each chunk are:
\begin{equation}
M_{\text{chunk}} = B M_{\text{token}} 
\end{equation}
The accumulated historical KV-cache size from the first to the $c$-th chunk is
\begin{equation}
M_{\text{hist}}(c) = c M_{\text{chunk}}
\end{equation}
MCF assigns each KV block a unique owner ($o_{i}$) according to the final sharding strategy, while allowing for replication across multiple nodes during the prefill stage. Let $r(c)$ denote the replica factor of the KV block generated at chunk $c$. Thus, $1 \leq r(c) \leq N$, where $r(c)=1$ corresponds to a fully sharded layout and $r(c)=N$ corresponds to full replication. MARS dynamically determines $r(c)$ based on two observations: 1) In the initial stage, when memory is abundant, early KV blocks are accessed repeatedly by subsequent chunks, thereby yielding the highest benefit from temporary replication (e.g., $r(0) = N$); 2) As the accumulated context lengthens and memory pressure intensifies, the system progressively curtails redundant replicas to alleviate resource contention. Furthermore, the distributed nature of KV cache facilitates more efficient scaling, thereby enhancing overall system scalability.

The chunk size $B$ determines both the computation granularity and the KV propagation granularity. A larger chunk improves accelerator utilization by increasing arithmetic intensity, but also increases the size of each KV block and the burst communication latency which is limited by hardware. Conversely, a smaller chunk reduces per-message latency but may provide insufficient computation to hide communication overhead. MARS therefore chooses $B$ by balancing compute-communication overlap and buffering capacity. The KV block produced by one chunk must fit within the available communication buffer:
\begin{equation}
B M_{\text{token}} \leq M_{\text{buf}},
\end{equation}
where $M_{\text{buf}}$ is the buffer capacity reserved for KV propagation. In addition, the computation time of a chunk should be large enough to overlap with the time required to disseminate its KV block, so that the communication overhead can be sufficiently hidden. The attention computation time of chunk $c$ can be approximated as
\begin{equation}
T_{\text{attn}}(B,c)
=
\frac{4 L H_q d \cdot B \cdot cB}{N P_{\text{node}}},
\end{equation}
where $P_{\text{node}}$ is the effective compute throughput of one node. The factor $4$ comes from the two dominant attention operations, $QK^T$ and $PV$, each contributing approximately two floating-point operations per multiply-add. The KV propagation time with replica factor $r_c$ is modeled as
\begin{equation}
T_{\text{prop}}(B,r_c)
=
\frac{M_{\text{chunk}} H_{\text{prop}}(r_c)}{BW_{\text{eff}}}
+
D_{\text{crit}}(r_c)\tau_{\text{link}},
\end{equation}
where $BW_{\text{eff}}$ is the effective interconnect bandwidth, $H_{\text{prop}}(r_c)$ is the distance-weighted cost of the selected propagation path, $D_{\text{crit}}(r_c)$ is the critical-path depth, and $\tau_{\text{link}}$ is the per-link latency. MARS selects the minimum feasible chunk size satisfying
\begin{align}\label{eq: min B}
B^\star &= \min B \nonumber \\
\quad
\text{s.t.}
& \quad
T_{\text{attn}}(B,c) \geq \gamma T_{\text{prop}}(B,r_c),\\
& \quad
B M_{\text{token}} \leq M_{\text{buf}} \nonumber
\end{align}
where $\gamma$ is an overlap safety factor which can be determined empirically. The choice of $B^\star$ favors the possible smallest chunk that can still hide KV propagation latency, avoiding excessive buffering and scheduling overhead.

Given a chunk size, MARS next determines the feasible replica factor under memory constraints. Let $S_{\text{node}}$ be the per-node memory budget available for KV cache. At chunk $c$, the average per-node KV memory footprint is
\begin{equation}
M_{\text{node}}(c)
=
\frac{M_{\text{chunk}}}{N}
\sum_{i=1}^{c} r(c)
\end{equation}
MARS enforces
\begin{equation}
M_{\text{node}}(c) \leq S_{\text{node}},
\end{equation}
which guarantees that temporary replication never exceeds the available KV-cache memory budget. If all historical KV blocks use the same replica factor at chunk $c$, a memory-bounded replica factor can be derived as
\begin{equation}
r_{\text{mem}}(c)
=
\left\lfloor
\frac{N S_{\text{node}}}{M_{\text{hist}}(c)}
\right\rfloor 
\end{equation}
The feasible replica factor is then
\begin{equation}
r_c
=
\max
\left(
1,
\min
\left(
N,
r_{\text{mem}}(c)
\right)
\right)
\end{equation}
This rule naturally creates a progressive convergence behavior. During early prefill, $M_{\text{hist}}(c)$ is small, so $r_c$ can be large and MCF can aggressively replicate KV blocks. As the context length increases, $M_{\text{hist}}(c)$ grows linearly, causing $r_c$ to decrease. Eventually, $r_c$ converges to one, producing the fully sharded layout required for memory-efficient decoding.

Memory capacity alone, however, is insufficient to decide which KV blocks should be replicated. A KV block generated early in prefill will be reused by many subsequent chunks, while a KV block generated near the end of prefill has limited future reuse. MARS therefore incorporates a reuse-aware criterion. Suppose there are $C$ kv blocks, for the KV block generated at chunk $c$, the number of future chunks that may reuse it is
\begin{equation}
R_c = C - c 
\end{equation}
Replicating this block is beneficial only if the future communication savings exceed the one-time replication cost. Let $h_{\text{copy}}$ be the average propagation cost of creating an additional replica, and let $h_{\text{save}}$ be the average propagation distance saved when future accesses are served locally or from a closer replica. The break-even condition is
\begin{equation}
(C-c)h_{\text{save}} > h_{\text{copy}} 
\end{equation}
This condition implies that earlier KV blocks should be replicated more aggressively than later KV blocks. MARS captures this effect using a reuse-bounded replica factor:
\begin{equation}
r_{\text{reuse}}(c)
=
\max
\left(
1,
\min
\left(
N,
1 + \left\lfloor \alpha (C-c) \right\rfloor
\right)
\right)
\end{equation}
where $\alpha$ controls the aggressiveness of reuse-based replication. The final replica factor is selected by jointly considering memory capacity and future reuse:
\begin{equation}
r(c)
=
\max
\left(
1,
\min
\left(
N,
r_{\text{mem}}(c),
r_{\text{reuse}}(c)
\right)
\right)
\end{equation}
This formulation biases the system toward replicating early, high-reuse KV blocks while keeping late KV blocks close to the final sharded layout.
MARS models the communication cost as the sum of two components: KV propagation cost and remote KV access cost. The propagation cost is paid when a newly generated KV block is replicated to additional nodes. For the KV block generated at chunk $c$, the distance-weighted propagation cost is
\begin{equation}
C_{\text{rep}}(c)
=
M_{\text{chunk}}
\left(
r(c)-1
\right)
h_{\text{dis}}(c)
\end{equation}
where $h_{\text{dis}}(i)$ represents the average physical cost of the mesh-aware propagation path. The total propagation traffic over the prefill stage is
\begin{equation}
C_{\text{rep,total}}
=
\sum_{c=1}^{C}
M_{\text{chunk}}
\left(
r(c)-1
\right)
h_{\text{dis}}(c)
\end{equation}
Remote KV access cost is incurred when a node needs a historical KV block that is not locally available. Assuming uniform replica placement, the number of nodes without a local replica of block $c$ is $N-r(c)$. Therefore, the distance-weighted access cost at chunk $c$ is
\begin{equation}
C_{\text{access}}(c) = \sum_{i=1}^{c} M_{\text{chunk}} (N - r(i)) h_{\text{near}}(i)
\end{equation}
where $h_{\text{near}}(i)$ is the average distance from a non-replica node to the nearest available replica. The total communication cost of MARS is
\begin{equation}
C_{\text{MARS}}
=
C_{\text{rep,total}} + \sum_{c=1}^{C} C_{\text{access}}(c)
\end{equation}
This expression exposes the trade-off controlled by MARS. Increasing the replica factor raises the first term because more replicas must be constructed, but reduces the second term because fewer future accesses need to traverse the network. In contrast, a ring-based KV streaming baseline has $r(c)=1$, resulting in no replica construction but repeated remote KV movement:
\begin{equation}
C_{\text{ring}}
=
M_{\text{chunk}}
(N-1)
h_{\text{ring}}
\frac{C(C+1)}{2}
\end{equation}
Full replication has $r(c)=N$, eliminating repeated remote accesses but requiring full KV replication:
\begin{equation}
C_{\text{full}}
=
M_{\text{chunk}}
(N-1)
h_{\text{dis}}
C
\end{equation}
MARS lies between these two extremes. It avoids the quadratic repeated communication pattern of ring-based execution while preventing the memory explosion of full replication.

The impact on TTFT can be modeled through the chunk-level execution time. Since TTFT is dominated by prefill latency, we approximate
\begin{equation}
T_{\text{TTFT}} \approx T_{\text{prefill}}
=
\sum_{c=1}^{C}
T_{\text{chunk}}(c)
\end{equation}
Each chunk consists of computation, communication, and synchronization. Because MCF overlaps KV propagation with local chunk computation, MARS models chunk latency as
\begin{equation}
T_{\text{chunk}}(c)
=
\max
\left(
T_{\text{comp}}(c)
T_{\text{comm}}(c)
\right)
+
T_{\text{sync}}(c)
\end{equation}
The communication latency is approximated by
\begin{equation}
T_{\text{comm}}(c)
=
\frac{
C_{\text{rep}}(c) + C_{\text{access}}(c)
}
{BW_{\text{eff}}}
+
D_{\text{crit}}(c)h_{\text{dis}}
\end{equation}
Here, $D_{\text{crit}}(c)$ captures the critical path of the selected propagation and access routes. For ring-based propagation, the critical path grows approximately linearly with $N$. For tree or butterfly-style propagation, the logical propagation depth grows as $O(\log N)$, although the realized latency on a physical mesh also depends on hop distance and link contention. Therefore, MARS selects mesh-aware paths that reduce both hop-weighted traffic and critical-path depth.

Overall, MARS solves the following constrained scheduling problem with constrains:
\begin{equation}
\min_{B, r_i(c), P_i(c)}
\sum_{c=1}^{C}
\max
\left(
T_{\text{comp}}(c),
T_{\text{comm}}(c)
\right),
\end{equation}
\begin{equation}
M_{\text{node}}(c)
\leq
S_{\text{node}},
\quad
\forall c,
\end{equation}
\begin{equation}
B M_{\text{token}} \leq M_{\text{buf}},
\end{equation}
\begin{equation}
r(C)=1
\end{equation}
The final constraint ensures that all temporary replicas created during prefill are removed before decoding begins, leaving a fully sharded KV-cache layout.

To enforce this convergence safely, MARS incorporates Progressive Replica Convergence as its runtime mechanism. Each temporary KV replica is associated with a lifecycle state. A valid replica can serve local attention computation. When the replica factor decreases, redundant replicas are marked as draining rather than immediately invalidated. A draining replica remains readable by in-flight chunks but is excluded from future scheduling decisions. Once all dependent chunks complete, the replica is invalidated and its buffer space is returned to the memory manager. This metadata-driven convergence avoids bulk KV migration because the owner shard is never removed. As a result, replica shrinking primarily involves directory updates and buffer-state transitions rather than additional data movement.

After convergence, MARS exposes a decode-ready layout: each KV block has a unique owner and the accumulated KV cache is evenly sharded across workers. Decode therefore uses a KV-stationary execution model. Each node attends over its local KV shard and emits compact online-softmax states, such as the local maximum, normalization sum, and partial output, which are reduced across nodes to obtain the global attention result. Thus, MARS bridges the layout mismatch between communication-optimized prefill and memory-scalable decode without requiring an explicit KV redistribution phase.
\section{Unified Progressive Execution Layout}
\label{sec:node}
\begin{figure}[htbp]
    \centering
    \includegraphics[width=\columnwidth]{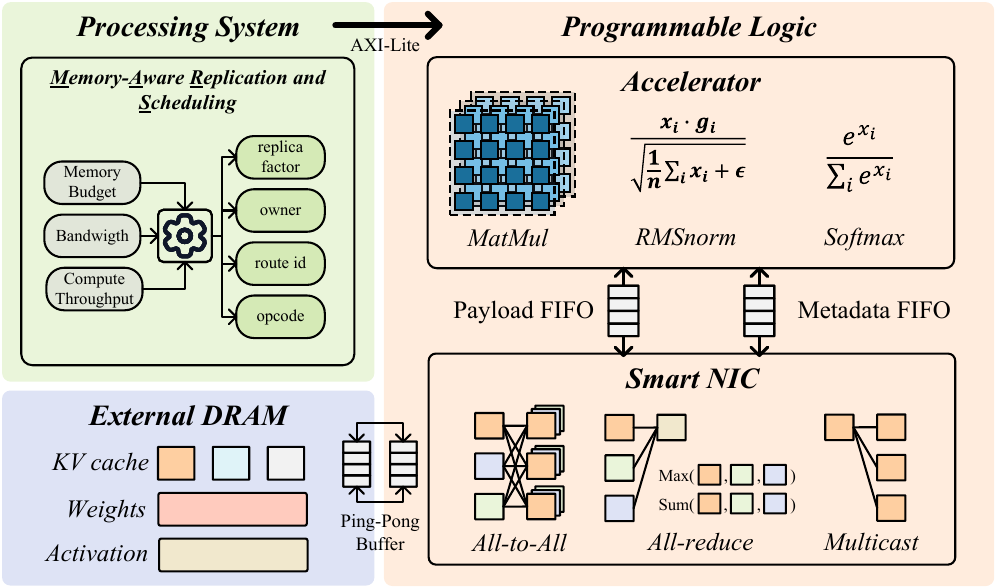}
    \caption{FPGA node architecture for Unified Progressive Execution Layout. The memory frontend streams local chunks through ping-pong buffers to sustain high-throughput computation. During prefill, the compute engine outputs KV-cache chunks; and for decoding stage, it produces partial attention states derived from computations over the local KV cache. MARS-generated metadata is attached to each payload and interpreted by the NIC datapath to perform collective operations.}
    \label{fig:fpga_node_arch}
\end{figure}
MARS produces a progressive KV-cache layout that changes across execution stages. To execute this dynamic layout efficiently, MCF introduces a Unified Progressive Execution Layout that maps MARS scheduling decisions into a FPGA-based cluster.

The central design principle is to avoid treating computation and communication as separate runtime phases. Instead, each FPGA node integrates a local compute engine with a SmartNIC. The compute engine produces either KV-cache blocks during prefill or compact partial attention states during decode. These outputs are packetized with MARS-generated metadata and sent directly to the NIC datapath through an AXI-Stream interface. The NIC then decodes the metadata and performs the required communication operation, such as multicast, all-to-all exchange, or all-reduce. In this way, MARS decisions are enforced directly in the data plane, bypassing host-side control flow latency and enabling full pipelining.

Figure~\ref{fig:fpga_node_arch} shows the architecture of a single FPGA node in MCF. Each node consists of a PS-side MARS runtime core and a PL-side compute-communication datapath. The MARS runtime core generates metadata, including \texttt{route\_id}, \texttt{opcode}, \texttt{KV id}, destination set, and collective type, and delivers it to the PL through AXI-Lite. This metadata later determines how each packet should be processed by the on-chip SmartNIC.

Inside the PL, a ping-pong DMA loader streams weights, activations, and local KV blocks from external DRAM through AXI-Full/AXI-MM. The loader uses double buffering to overlap burst memory reads with kernel execution: while one buffer feeds the GEMM/attention kernel, the other is refilled from DRAM. This design hides DRAM latency and improves memory bandwidth utilization.

The GEMM kernel performs QKV projection and local attention computation, then pushes computed payloads into a packetization buffer. The buffer maintains separate payload and metadata FIFOs and emits AXI-Stream packets to the SmartNIC. During prefill, the payload contains KV-cache blocks for MARS-controlled propagation. During decode, it contains compact partial attention states such as local maximum, exponential sum, and partial output. The SmartNIC decodes the metadata and performs the corresponding operation, including forwarding, multicast, all-to-all, all-reduce, or max/sum/merge reduction. By decoupling the compute kernel from NIC communication, MCF allows computation and communication to proceed asynchronously and overlap in time.

The compute engine performs the local transformer computation required by both prefill and decode. During prefill, the core consumes the current local chunk, computes QKV projections, and writes the generated KV block into the local KV buffer as its owner shard. If MARS determines that the block should be temporarily replicated, the same KV block is also emitted to the packetization buffer as a communication payload. During local attention computation, the core reads locally available KV blocks, including both owner shards and valid temporary replicas, according to the local replica directory. During decode, after MARS has converged the KV cache to a uniform sharded layout, the core computes attention over the local KV shard and emits compact partial softmax states, such as the local maximum, local exponential sum, and partial output vector. These states are later reduced across nodes to form the exact global attention result.

The output of the compute engine is decoupled from the NIC by a packetization buffer. This buffer contains a payload FIFO and a metadata FIFO. The payload FIFO stores activation tensors, including prefill KV blocks and decode partial states. The metadata FIFO stores the control information generated from MARS, including routing identifiers, opcode, collective type, destination set, KV block identifier, and reduction operator. The packetizer combines the payload and metadata into an AXI-Stream packet. This design allows the compute pipeline to continue producing data even when the NIC experiences temporary congestion, as long as the packetization buffer has sufficient capacity.
Each outgoing packet follows a unified format: $\text{Packet} = \{\text{Meta}, \text{Payload}\}$, where the Meta describes how the payload should be processed by the NIC: \text{Meta} = \{\text{stage}, \text{op}, \text{src}, \text{dst}, \text{kv\_id}, \text{route\_id}, \text{collective}, \text{reduce\_op}\}.

The \texttt{stage} field distinguishes prefill and decode. The \texttt{op} field specifies whether the packet carries a KV replica, a replica invalidation request, a query or state packet, or a reduction message. The \texttt{src}, \texttt{dst}, and \texttt{route\_id} fields identify the routing action selected by MARS. The \texttt{collective} field indicates whether the packet participates in point-to-point forwarding, multicast dissemination, all-to-all exchange, or all-reduce. The \texttt{reduce\_op} field specifies the operation required during decode, such as maximum comparison, summation, or online-softmax state merge.

The NIC datapath is metadata-driven. When an AXI-Stream packet arrives from the packetization buffer, the packet parser separates metadata from payload. The metadata decoder then dispatches the packet to the appropriate hardware engine. KV replication packets are sent to the forwarding or multicast engine, which follows the mesh-aware route generated by MARS. Collective packets are handled by the all-to-all engine. Decode partial-state packets are sent to the reduction engine, which performs the maximum and summation operations required to merge distributed softmax statistics. Since the NIC performs these operations directly in hardware, the host CPU is removed from the critical path of KV routing and decode-state reduction.

This unified packetized interface supports both stages of inference. During prefill, the payload is dominated by KV-cache data. MARS may assign a high replica factor to early high-reuse KV blocks, and the corresponding metadata instructs the NIC to multicast these KV blocks through a mesh-aware dissemination path. When memory pressure increases, MARS reduces the replica factor and generates metadata that marks redundant replicas as draining or invalid. Since the owner shard is never removed, replica convergence only requires directory updates and buffer-state transitions, avoiding bulk KV migration.

During decode, the same datapath switches to a KV stationary execution mode. The accumulated KV cache is already evenly sharded across nodes by MARS, so the system avoids moving or replicating bulk KV data. Instead, each node computes attention over its local KV shard and exchanges only lightweight query or partial-state packets. The reduction engine merges local softmax states using online-softmax-compatible operations. This behavior is similar to distributed Flash-Decoding-style execution: the long KV dimension is partitioned across workers, local attention is computed independently, and compact partial states are reduced to obtain the final attention output.

The benefit of this design is that prefill and decode share the same hardware communication substrate while using different metadata policies. Prefill uses KV-centric dissemination to exploit early memory headroom and reduce repeated remote KV access. Decode uses query/state-centric routing over a stationary, balanced KV layout. The transition between the two does not require a separate data redistribution phase because MARS guarantees that all temporary replicas have converged to owner shards by the end of prefill.

Overall, Unified Progressive Execution Layout turns MARS from an analytical scheduling algorithm into an executable hardware protocol. MARS determines the layout, route, and operation type; the compute core produces the corresponding payload; the packetization buffer attaches metadata; and the NIC datapath executes the required communication or reduction operation. By fusing local computation with metadata-driven communication, MCF supports communication-efficient chunked prefill and distributed Flash-Decoding-style decode within a single FPGA execution framework.
\section{Evaluation}
\subsection{Experimental Setup}
We evaluate MCF on a cluster of AMD-Xilinx KR260 FPGA boards. Each board contains
a Zynq UltraScale+ MPSoC with programmable logic and external DDR memory. The
FPGA fabric implements the transformer compute core, packetization buffer, and
metadata-driven NIC datapath described in Section~\ref{sec:node}. The ARM cores
are used only for initialization and for running the MARS control logic; they
are not involved in critical-path tensor movement.
We adopt Qwen2.5-1.5B as our benchmark model: it follows a standard 
decoder-only Transformer architecture and, despite its compact size, exhibits 
the prefill and decode behavior representative of modern long-context LLMs. 
Unless otherwise specified, the model weights and KV-cache entries are 
quantized to INT4 in the fixed-point format natively supported by our FPGA 
compute core. We sweep the input sequence length from short to long contexts 
and evaluate several prefill chunk sizes to characterize the interplay among 
compute granularity, communication overhead, and on-chip memory usage.

We compare MCF against three baselines. \textbf{Ring Attention}~\cite{liu2024ringattention}
streams KV blocks across nodes in a ring-like pattern during prefill.
\textbf{Zigzag Attention}~\cite{liu2025zigzagattention} uses an alternative
sequence layout to reduce causal attention load imbalance. \textbf{PD
disaggregation}~\cite{zhong2024distserve} separates the prefill and decode
stages across different compute instances to improve system-level resource
utilization. We compare against dataflow-equivalent implementations of these
baselines. Each baseline follows the KV placement and communication pattern
proposed in the corresponding work while using the same compute core and
communication interface. MCF dynamically chooses replica factors using MARS and
progressively converges temporary replicas to the final sharded layout before
decode.
\begin{figure}[t]
  \centering
  \includegraphics[width=\linewidth]{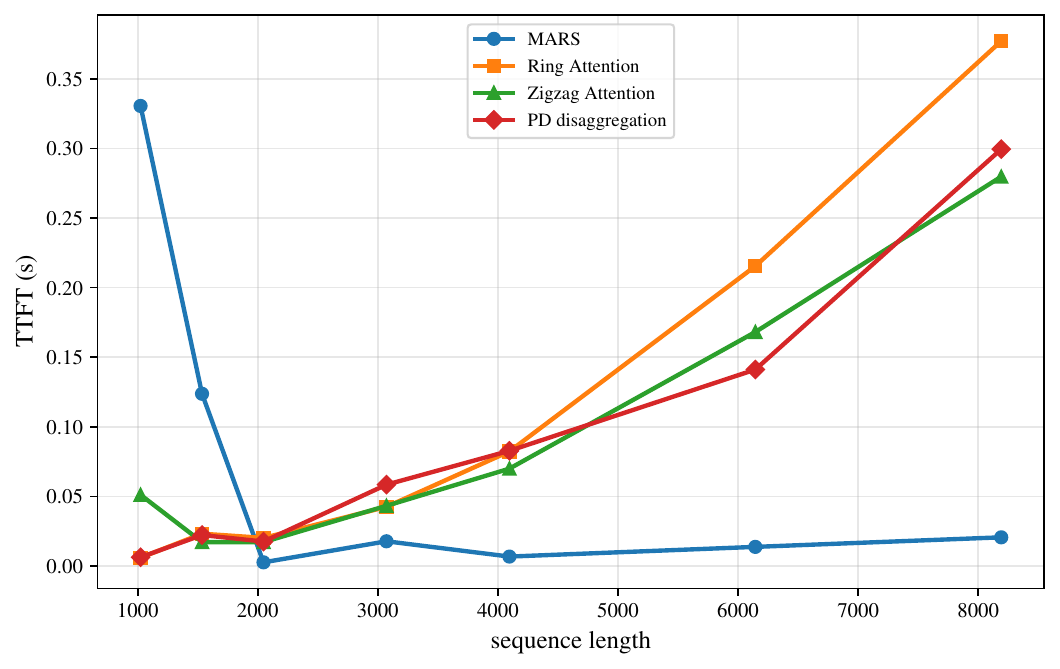}
  \caption{Prefill TTFT comparison up to 8K sequence length. MCF achieves lower
  TTFT than Ring Attention, Zigzag Attention, and PD disaggregation by using
  temporary KV replication and scheduled remote KV movement during prefill.}
  \label{fig:res}
\end{figure}
\subsection{End-to-End Prefill Latency}
Figure~\ref{fig:res} reports the prefill TTFT as the input sequence length increases up to 8K. MCF achieves the lowest TTFT for medium and long contexts and shows a much flatter latency curve than the baselines as the sequence length increases. Ring Attention, Zigzag Attention and PD disaggregation exhibit rapidly increasing TTFT because longer prompts require more historical KV blocks to be streamed or handed off during prefill. By pairing \textit{heavy} blocks at the tail of the sequence (i.e., those  attending to many preceding tokens under the causal mask) with \textit{light} blocks at the head, Zigzag Attention exploits the asymmetric workload inherent to causal LLMs and thereby achieves a moderate performance gain. In contrast, MCF reduces the exposed latency of remote KV access by using temporary replication and scheduled KV movement. For short sequences, the benefit of MCF is less pronounced and may even be partially offset by scheduling and replication overhead. This is because the number of chunks is small and the fixed chunk/block granularity does not provide enough reuse opportunities to amortize the cost of replica creation. Meanwhile, memory pressure is low in the early prefill stage, so MARS tends to assign a larger replica factor to early KV blocks. These additional temporary replicas increase communication and metadata-processing overhead, which can make MCF less efficient for very short prompts. As the sequence length grows, however, the replicated KV blocks are reused by more subsequent chunks, and the replication overhead is amortized over a larger amount of attention computation.

The improvement of MCF comes from two properties. First, MARS creates temporary KV replicas for high-reuse blocks in the early prefill stage, allowing later chunks to access a larger fraction of historical KV blocks locally or through a shorter pre-scheduled datapath. Second, because the required historical KV blocks are known before each chunk is computed, MCF can schedule remote KV movement ahead of use instead of handling it as reactive demand fetch. This reduces the communication exposed on the prefill critical path. Unlike Ring Attention and Zigzag Attention, which repeatedly stream historical KV blocks during attention computation, MCF trades temporary memory usage and modest additional communication for lower visible remote-KV access latency in the long-context regime.
\begin{figure}[t]
    \centering
    \includegraphics[width=\linewidth]{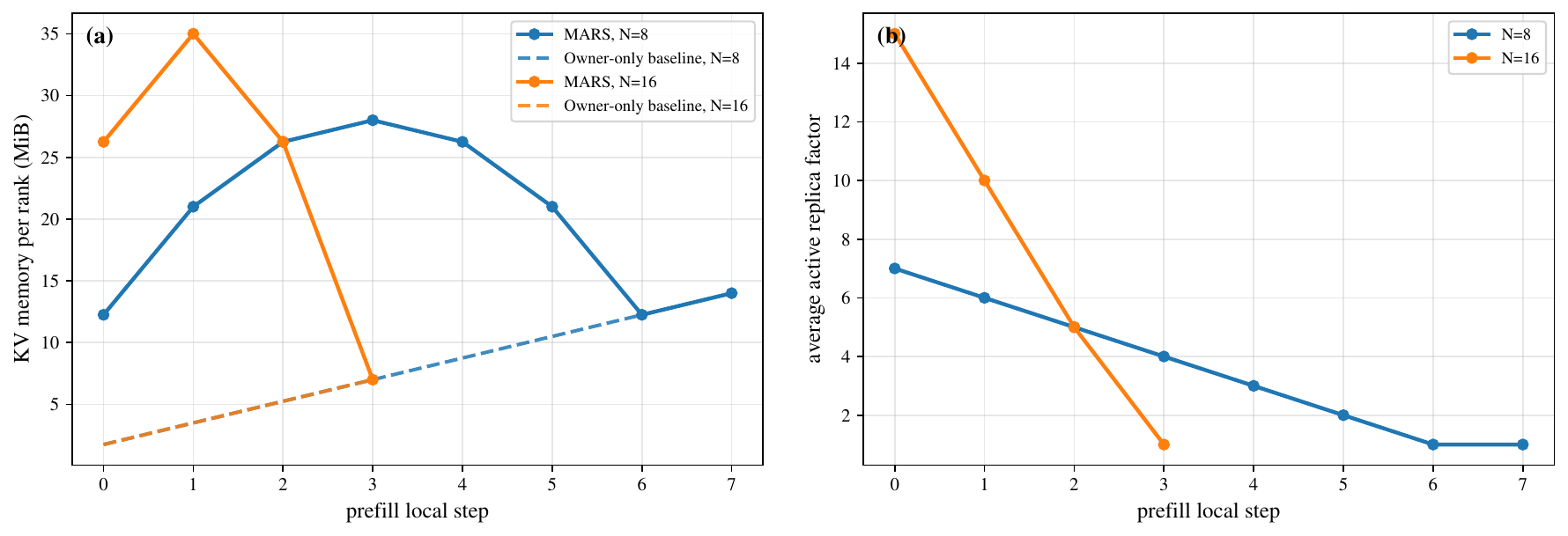}
    \caption{Memory behavior of MARS during 16K prefill for $N=8$ and $N=16$.
    Subfigure~(a) compares the actual per-rank KV memory usage with the
    theoretical owner-only baseline without temporary replicas. Subfigure~(b)
    shows the average active replica factor over prefill steps. MARS temporarily
    increases KV memory during early prefill and converges to the owner-only
    layout before decode.}
    \label{fig:mars_memory_trace}
\end{figure}

\subsection{Memory Usage and Layout Convergence}
Figure~\ref{fig:mars_memory_trace} shows the memory behavior of MARS under 16K
prefill for $N=8$ and $N=16$ logical nodes. The left subfigure compares the
actual per-rank KV memory usage with an owner-only baseline, where each rank
stores only its locally owned KV blocks and no temporary replicas are created.
MARS uses more memory than this baseline during the early and middle prefill
stages because it intentionally replicates high-reuse KV blocks. However, the
memory footprint decreases as prefill progresses and eventually converges to
the owner-only layout.

The right subfigure explains this behavior by showing the average active replica
factor. At the beginning of prefill, early KV blocks have the highest future
reuse, so MARS assigns a larger replica factor. As more chunks are processed,
memory pressure increases and the scheduler gradually reduces the replica
factor. At the end of prefill, all temporary replicas are drained, and only the
owner KV blocks remain. This confirms that MCF does not introduce permanent KV
replication and that the decode stage starts from a KV-stationary, owner-only
sharded layout.

\begin{figure}[t]
    \centering
    \includegraphics[width=\linewidth]{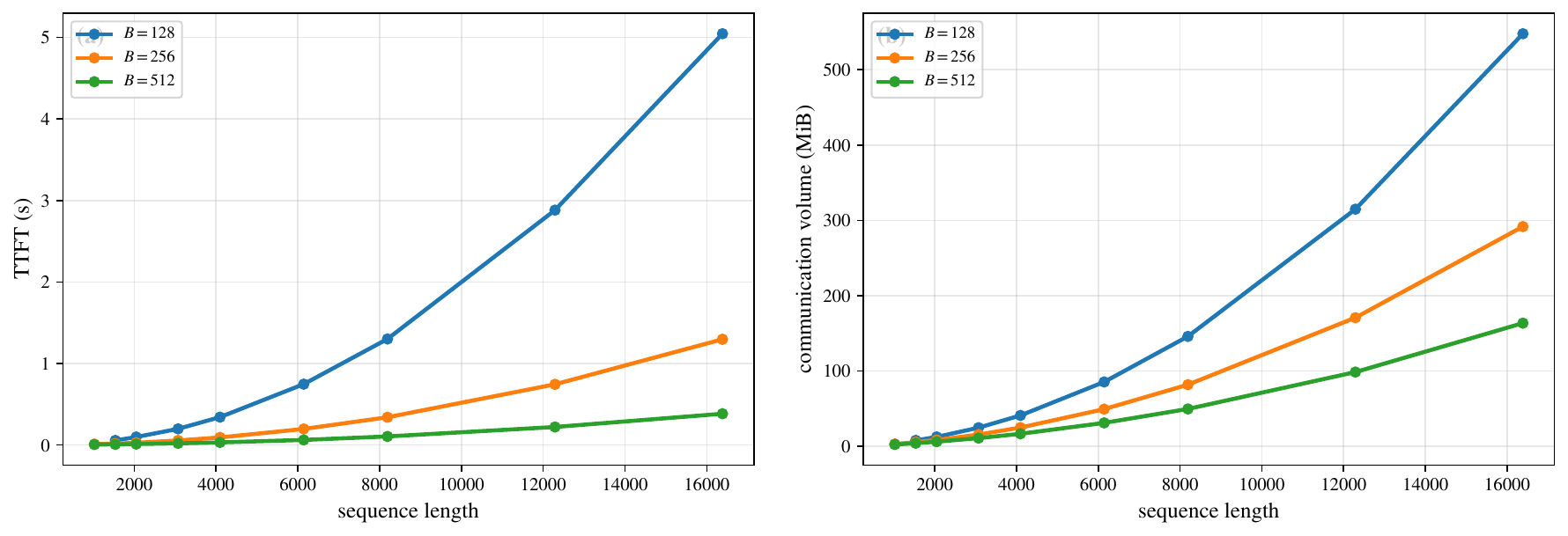}
    \caption{Impact of prefill chunk size on MARS. Subfigure~(a) shows the TTFT
    under different chunk sizes, while Subfigure~(b) shows the corresponding
    communication volume. Larger chunks reduce the number of chunk-processing
    rounds and metadata-guided KV movements, leading to lower TTFT and lower
    total communication volume.}
    \label{fig:mars_chunksize}
\end{figure}
\subsection{Impact of Chunk Size}
Figure~\ref{fig:mars_chunksize} evaluates how the prefill chunk size affects
MARS. Larger chunk sizes provide lower TTFT and lower communication volume
across the tested sequence lengths. This is because smaller chunks increase the
number of prefill rounds. More rounds lead to more frequent scheduling
decisions, more metadata processing, more replica updates, and more remote KV
movements. Larger chunks amortize these overheads over more tokens and reduce
the number of inter-chunk KV accesses.

This result highlights an important trade-off in chunked prefill. Smaller chunks
offer finer scheduling granularity and can reduce per-chunk working-set size,
but they also increase communication and scheduling overhead. Larger chunks
reduce these overheads, but may reduce scheduling flexibility and increase the
latency of each individual chunk. In our evaluated range, larger chunks provide
better performance because the reduction in communication and scheduling
overhead dominates.

\subsection{Scalability with Number of Nodes}

Table~\ref{tab:node_scaling_ttft_memory} reports 16K prefill performance under
different logical node counts. MARS reduces TTFT as the number of nodes
increases, from $1.68$s at two nodes to $0.62$s at sixteen nodes. This shows
that MCF benefits from distributing prefill computation and KV storage across
more nodes. The scaling is not perfectly linear because communication,
synchronization, and replica-management overheads increase with the node count.

MCF achieves lower TTFT than Ring Attention and Zigzag Attention at all evaluated
node counts. At sixteen nodes, MCF reaches $0.62$s, while Ring Attention and
Zigzag Attention require $1.91$s and $1.81$s, respectively. This corresponds to
approximately $3.1\times$ speedup over Ring Attention and $2.9\times$ speedup
over Zigzag Attention. The communication volume of MCF is slightly higher than
the streaming baselines at larger node counts. For example, at sixteen nodes,
MCF transfers $5.12$GiB, while Ring Attention and Zigzag Attention transfer
about $4.40$GiB. This is consistent with the design of MARS: MCF does not aim to
minimize total transferred bytes. Instead, it trades additional replica-creation
and prefetch traffic for lower exposed communication latency on the prefill
critical path.

\begin{table}[h]
\centering
\caption{TTFT and communication volume for 16K prefill across different logical
node counts. MCF achieves lower TTFT than the streaming baselines while using
slightly higher communication volume due to temporary replica creation and
scheduled KV prefetch.}
\label{tab:node_scaling_ttft_memory}
\begin{tabular}{llrr}
\toprule
Scheme & Nodes & TTFT (s) & Comm (GiB) \\
\midrule
MARS & 2 & 1.68 & 2.13 \\
 & 4 & 1.16 & 3.02 \\
 & 8 & 0.97 & 3.71 \\
 & 16 & 0.62 & 5.12 \\
\midrule
Ring Attention & 2 & 3.96 & 1.94 \\
 & 4 & 2.46 & 2.99 \\
 & 8 & 2.29 & 3.70 \\
 & 16 & 1.91 & 4.40 \\
\midrule
Zigzag Attention & 2 & 3.81 & 1.94 \\
 & 4 & 2.35 & 2.99 \\
 & 8 & 2.18 & 3.69 \\
 & 16 & 1.81 & 4.40 \\
\bottomrule
\end{tabular}
\end{table}

These results show that MCF improves long-context prefill latency by
using temporary memory headroom to place high-reuse KV blocks closer to future
consumers. The temporary replicas are progressively removed, so the final KV
layout remains owner-only sharded and decode-friendly.
\subsection{Decoding Performance and Power Report}
\begin{table*}[t]
\centering
\caption{Performance and Metrics Comparison of DDR-Based FPGA LLM Accelerators}
\label{tab:fpga_comparison}
\renewcommand{\arraystretch}{1.15}
\setlength{\tabcolsep}{3pt}
\resizebox{\textwidth}{!}{%
\begin{tabular}{l|ccc|cccccccc|c}
\toprule
 & \multicolumn{3}{c|}{\textbf{DDR-Based Cloud FPGA}} 
 & \multicolumn{8}{c|}{\textbf{DDR-Based Embedded FPGA}} 
 & \textbf{Ours} \\
\cmidrule(lr){2-4} \cmidrule(lr){5-12} \cmidrule(lr){13-13}
 & ChatOPU & LightMamba & \textit{Hummingbird} 
 & Li~\textit{et al.} & LlamaF & \textit{Hummingbird} & TeLLMe & MEADOW & SECDA-LLM & F-BFQ & OD-Qwen 
 & \textbf{MCF}$^{\dagger}$ \\
 & \cite{chatopu} & \cite{lightmamba} & \cite{hummingbird}
 & \cite{li2025pushing} & \cite{llamaf} & \cite{hummingbird} & \cite{tellme} & \cite{meadow} & \cite{secda} & \cite{fbfq} & \cite{odqwen}
 & (4 Nodes)\\
\midrule
Device     & U200    & VCK190  & U250    & KV260   & ZCU102 & ZCU104/KV260 & KV260 & ZCU102 & PYNQ-Z1 & KV260 & KV260   & \textbf{KR260} \\
BW(GB/s)       & 19.2    & 12/76.8 & 76.8    & 19.2    & 19.2   & 34.1/19.2    & 19.2  & $\le$19.2 & 0.4 & 19.2  & 19.2    & \textbf{19.2}    \\
\midrule
LUT        & 755K    & 107K    & 168K    & 78K     & 164K   & 66K/26K      & --    & --     & --      & 81\%* & --      & \textbf{82K}$^{\ddagger}$    \\
FF         & 998K    & 130K    & 218K    & 105K    & 171K   & 92K/35K      & --    & --     & --      & --    & --      & \textbf{88K}$^{\ddagger}$    \\
DSP        & 1091    & 228     & 771     & 291     & 528    & 362/179      & --    & --     & --      & 14\%* & --      & \textbf{127}$^{\ddagger}$    \\
B/URAM     & 417/540 & 912/61  & 361/76  & 37/10   & 223/-- & 133/40, 59/18& --    & --     & --      & 81\%* & --      & \textbf{129/20}$^{\ddagger}$ \\
MHz        & 300     & 400     & 300     & 300     & 205    & 266/300      & 250   & --     & --      & --    & 300     & \textbf{200}    \\
W          & --      & 3.2     & 14.61   & 6.57    & 5.08   & 7.09/3.81    & $<$5  & $<$10  & --      & --    & --      & \textbf{4.52}$^{\ddagger}$    \\
\midrule
Model      & opt-350m & mamba2-2.7b & llama3-8b & llama2-7b & llama-1.1b & llama3-8b & BitNet-0.7b & OPT-1.3b & TinyLlama-1.1b & TinyLlama-1.1b & Qwen2.5-0.5b & \textbf{Qwen2.5-1.5B} \\
Quant.     & SparseW16 & W4         & W4        & W4        & W8         & W4        & W1.58A8     & INT8     & BFP-Q3\_K      & BFP-Q2/Q3\_K   & W4 (AWQ)     & \textbf{W4} \\
Token/s    & 43.2      & 7.21       & 19.4      & 4.9       & 1.48       & 8.6/4.8   & 25          & 2.0      & 0.58           & 5.2            & 5.1          & \textbf{17.3}$^{\dagger}$           \\
PW. Eff.   & --        & 0.87       & 1.51      & 0.74      & 0.09       & 1.39/1.44 & $\sim$5.0   & $\sim$0.20 & --           & --             & --           & \textbf{0.81}           \\
\bottomrule
\end{tabular}%
}
\\
\footnotesize
\raggedright
$^{\dagger}$MCF results are obtained on a cluster of 4 KR260 nodes. Token/s denotes the end-to-end throughput of the cluster.\\
$^{\ddagger}$Resource utilization and power are reported per single KR260 node; all 4 nodes share an identical bitstream and configuration.\\
$^{*}$F-BFQ reports resource utilization as a percentage of the KV260 board capacity rather than absolute counts.\\
\end{table*}

Table~\ref{tab:fpga_comparison} compares MCF with existing DDR-based FPGA LLM
accelerators. MCF achieves $17.3$ token/s on Qwen2.5-1.5B using a cluster of
four KR260 nodes. Compared with prior embedded FPGA accelerators, MCF supports a
larger model than several KV260/ZCU102-based systems and provides competitive
end-to-end decoding throughput. This result demonstrates that the proposed
KV-stationary decode path can effectively use multiple low-power FPGA nodes for
distributed LLM inference.

The power efficiency of MCF is not the highest among the listed systems. This is
expected because MCF integrates not only the compute engine, but
also the packetization buffer, and NIC-side communication datapath required for distributed inference. These modules increase LUT/BRAM usage and consume additional power compared with
single-board accelerators that only optimize local model execution. In addition,
MCF exchanges query vectors, softmax statistics, and partial attention states
during decode, so part of the system power is spent on communication rather than
pure matrix computation. As a result, MCF trades some single-board power
efficiency for the ability to scale KV-cache storage and decode computation
across multiple FPGA nodes.


Overall, Table~\ref{tab:fpga_comparison} shows that MCF provides competitive decode throughput on a distributed embedded-FPGA platform while supporting communication-aware KV-stationary execution. Although its power efficiency is lower than highly optimized single-node designs, MCF targets a different design
point: scalable long-context inference with distributed KV storage and communication integrated into the FPGA datapath.
\section{Related Work}
We discuss related work for context parallelism algorithm and dynamic memory management for LLM inference, and then draw the difference between state of the art approach and MCF. 
\subsection{Context Parallelism Algorithms}
Context parallelism partitions maximum sequence dimension across devices. For example, Li et al.~\cite{li2023sequenceparallelism} proposed Ring Self-Attention (RSA) in which each device stores only a subsequence, rather than the full sequence. RSA circulated key embeddings across devices to compute attention scores. After that, it circulates value embeddings to compute the attention outputs. This way, RSA improved scalability for long-sequence training to some extent. However, RSA still required the distributed attention-score blocks to be explicitly stored in memory.

To avoid storing explicitly attention-score blocks in memory, Liu et al.~\cite{liu2024ringattention} proposed Ring Attention, which implemented ring-based sequence parallelism by combining KV-block circulation with blockwise Transformer computation. In Ring Attention, the input sequence is partitioned across devices, so that each device keeps a local Q, K, and V block initially. Then, KV blocks circulate based on a ring structure while each query block fixed to an associated device. 

As each KV block arrives, the device performs partial attention on these KVcache and update the attention score in a flashattention manner~\cite{dao2022flashattetion}. This reduces the peak activation memory per device to depend primarily on the local block size rather than the full sequence length. 
However, existing ring-based methods still impose a static communication structure. RSA and Ring Attention assume that KV blocks traverse a predetermined ring. This is efficient when devices are homogeneous and communication can be fully overlapped with blockwise computation. In practice, long-context inference can be more dynamic: memory pressure changes across prefill and decoding; different context blocks have different reuse patterns; and bandwidth, latency, or device availability can vary across the system. A fixed ring traversal can therefore become inefficient when some links are slower or when certain KV blocks would benefit from temporary replication rather than repeated circulation.

ZigZag Attention\cite{liu2025zigzagattention} addresses this bottleneck by reordering the sequence blocks to balance the workload. Instead of a linear traversal, it employs a "zigzag" scheduling strategy that pairs computation-heavy blocks (near the end of the sequence) with computation-light blocks (near the beginning) within the same communication step. By ensuring that the workload is approximately equal across all devices in the ring, ZigZag Attention achieves nearly $2\times$ higher computational efficiency compared to unoptimized causal Ring Attention, all while maintaining the same communication overhead and memory benefits.
While ZigZag optimizes the load balance within a static ring, it does not address the fundamental resource tension between the two distinct phases of long-context inference. The Prefill phase is highly compute-bound, requiring massive throughput to process the prompt, whereas the Decoding phase is strictly memory-bandwidth bound (IO-bound) and sensitive to per-token latency. Attempting to manage both within a single, fixed ring often leads to "prefill-stalls," where long context processing blocks the generation of shorter requests.
Prefill-Decoding Disaggregation\cite{zhong2024distserve} resolves this by decoupling the two phases into distinct hardware pools: prefill nodes use TP/PP/CP for high-throughput ingestion on premium GPUs, while decoding nodes specialize in autoregressive generation. KV cache is handed off between pools over high-speed interconnects. This separation enables independent scaling and optimizes TCO.


The proposed MCF addresses this limitation by treating context parallelism not merely as a static sequence-partitioning method, but as a dynamic memory-communication scheduling problem. Instead of forcing every KV block to follow the same fixed traversal pattern, MCF dynamically decides whether a context block should be sharded, replicated, migrated, or circulated according to the current inference stage, memory pressure, expected reuse, and communication cost. In this sense, MCF generalises ring-based context parallelism: it preserves the benefit of distributing long-context attention across devices, but removes the assumption that one fixed ring schedule is always optimal.
\subsection{Dynamic Memory Management for LLM Inference}

Efficient LLM serving requires careful management of model weights, intermediate activations, attention buffers, and the KV cache. For long-context inference, the KV cache is particularly important because it grows with the number of processed tokens and must be repeatedly accessed during decoding. Kwon et al.~\cite{kwon2023vllm} proposed PagedAttention, which treats KV-cache storage using a paging-inspired mechanism, reducing memory fragmentation and enabling flexible KV-cache sharing across requests. Sheng et al.~\cite{sheng2023flexgen} proposed FlexGen for high-throughput generation by coordinating GPU, CPU, and disk memory. Such related work discusses that LLM inference performance is strongly constrained by memory placement and scheduling.

Agrawal et al.~\cite{agrawal2024sarathi}  introduces chunked prefill and stall-free scheduling. DistServe and Splitwise separate prefill and decoding across different GPU resources ~\cite{zhong2024distserve,patel2024splitwise}.

Other work focuses directly on KV-cache movement and reuse. Liu et al.~\cite{liu2024cachegen} compressed and streamed KV caches to reduce context-loading delay when reusing long contexts across requests. Qin et al.~~\cite{qin2025mooncake} proposed a KV-cache-centric serving architecture by separating prefill and decoding clusters and using CPU, DRAM, SSD, and NIC resources to build a distributed KV-cache pool. These systems recognised that KV-cache movement could be a main bottleneck in LLM serving. 
Unlike these systems, which focus on KV reuse and memory hierarchy, MCF introduces stage-aware KV placement inside CP itself: MARS temporarily replicates high-reuse blocks during low-pressure prefill stages and falls back to sharding as memory tightens
MCF exploits these stage-dependent changes through MARS: high-reuse blocks can be temporarily replicated when memory pressure is low, while the system can fall back to sharding when memory becomes scarce. Thus, MCF introduces a dynamic memory-management layer inside CP, rather than treating KV-cache layout as fixed once the sequence has been partitioned.
\section{Conclusion}
We propose MARS, a stage-aware KV-cache orchestrator, that dynamically replicates, removes, and converges KV blocks across prefill stages. MARS improves access locality for highly reused historical KV blocks while ensuring that temporary replication remains memory-bounded and that the final layout is compatible with efficient sharded decoding. We further introduced a distributed chunked-prefill execution model that pipelines local attention computation with KV propagation and remote KV-cache fetching, allowing communication latency to be hidden behind useful computation whenever feasible. For example, prior work did not explicitly identify the runtime conditions under which communication overhead could be hidden, as formulated in Equation~(\ref{eq: min B}). We also propose a hardware-software co-designed execution framework that connects MARS-level scheduling decisions to the underlying distributed execution substrate. By using a metadata-driven FPGA compute-NIC datapath, the framework supports dynamic KV routing, replication, and collective operations in hardware, reducing software overhead and enabling high-performance distributed inference.

Overall, our work shifts distributed long-context LLM inference from static-level KV orchestration framework to dynamic-level memory- and computing-aware KV-cache orchestration framework. Therefore, we believe that our framework will provide a practical initial road map for improving long-context TTFT under dynamically varying interconnect bandwidth and memory constraints while preserving a decode-ready KV layout for subsequent generation.


\bibliographystyle{ACM-Reference-Format}
\bibliography{sample-base}

\appendix









\end{document}